\documentclass[%
 reprint,
 superscriptaddress,
 amsmath,amssymb,
 aps,
 prl,
]{revtex4-2}

\usepackage{graphicx}
\usepackage{dcolumn}
\usepackage{bm}

\usepackage{xcolor}

\begin{document}

\preprint{APS/123-QED}

\title{Spontaneous rotation and propulsion of suspended capsules in active nematics}


\author{Júlio P. A. Santos}
 \affiliation{Centro de Física Teórica e Computacional, Faculdade de Ciências, Universidade de Lisboa, 1749-016 Lisboa, Portugal.}
 \affiliation{Departamento de Física, Faculdade de Ciências, Universidade de Lisboa, P-1749-016 Lisboa, Portugal.}
 \affiliation{University of Vienna, Faculty of Physics, Kolingasse 14-16, 1090 Vienna, Austria}

\author{Margarida M. Telo da Gama}%
 \affiliation{Centro de Física Teórica e Computacional, Faculdade de Ciências, Universidade de Lisboa, 1749-016 Lisboa, Portugal.}
 \affiliation{Departamento de Física, Faculdade de Ciências, Universidade de Lisboa, P-1749-016 Lisboa, Portugal.}
\affiliation{International Institute for Sustainability with Knotted Chiral Meta Matter, Hiroshima University, Higashihiroshima 739-8511, Japan.}

\author{Rodrigo C. V. Coelho}
 \email{rcvcoelho@cbpf.br}
 \affiliation{Centro de Física Teórica e Computacional, Faculdade de Ciências, Universidade de Lisboa, 1749-016 Lisboa, Portugal.}
 \affiliation{Departamento de Física, Faculdade de Ciências, Universidade de Lisboa, P-1749-016 Lisboa, Portugal.}
 \affiliation{Centro Brasileiro de Pesquisas Físicas, Rua Xavier Sigaud 150, 22290-180 Rio de Janeiro, Brazil}

\date{\today}

\begin{abstract}
We investigate the dynamics of elastic capsules suspended in two-dimensional active nematic fluids using lattice Boltzmann simulations. The capsules, modeled as flexible membranes enclosing active internal regions, exhibit a rich variety of behaviors shaped by their geometry and the interplay between internal and external activity. Circular capsules with active interiors undergo persistent rotation driven by internally confined +1/2 topological defects. Axisymmetric capsules, such as boomerangs, develop directed motion along their axis of symmetry due to unbalanced active forces generated by defect distributions near their boundaries. We further show that capsule flexibility suppresses motility and rotation, as active stresses are dissipated into shape deformations. These findings reveal how shape, deformability, and defect dynamics cooperate to produce emergent motility in soft active matter, with potential applications in the design of microswimmers and drug delivery vehicles.
\end{abstract}

\maketitle


Elastic thin--shells, deformable surfaces that sustain both bending and stretching, play a central role across soft matter physics and biology, from viral capsids and synthetic vesicles to red blood cells and tissue spheroids~\cite{Lin_2018, Flexible1984, Shell2015}. When these shells are immersed in active nematic fluids, which are nonequilibrium media of energy-consuming, orientationally ordered constituents, they are subjected to stresses that drive complex and often unexpected dynamics. The problem becomes even richer when the shell encloses a distinct internal phase, passive or active, introducing new couplings between internal and external flows, active stresses, and mechanical resistance.

Active nematics exert anisotropic and time-dependent stresses on the shell through self-organized flows and the motion of topological defects~\cite{Hydrodynamics-AN-Review,Active-nematics-Review,AN-Ramaswamy-Review,Geo-Top-AN,Orient-order-defect-AN,topo-dynamics-AN-vesicles}. The response of the capsule is further modulated by its interior: a passive fluid sets the balance between external stresses and internal pressure, while an active interior generates its own stresses~\cite{shape-dynamics-A-vesicles,topo-morfo-A-shells}. Even a passive interior is far from trivial: the chaotic dynamics of active nematics~\cite{AditiSimha2002, Narayan2007, Sanchez2012, Zhou2014, Duclos2018, Li2018, Kumar2018, Tan2019} enhance the dispersion and drive the anomalous transport of tracers and droplets~\cite{tracer-ActTurb, gotas}. Symmetry-breaking effects are also prominent: chiral asymmetric inclusions rotate persistently~\cite{engrenagem-AN, Houston_2023}, akin from observed in bacterial suspensions~\cite{Sokolov2009, DiLeonardo2010}, while symmetric ones typically rotate only randomly~\cite{engrenagem-AN}, unless confinement stabilizes vortical flows~\cite{Ye2024}. On the other hand, circular cavities with appropriate size filled with active nematics can trap a pair of rotating defects in a yin–yang–like configuration~\cite{capsulas-exp, rotating-defects}, recently extended to three dimensions~\cite{PhysRevX.14.041002}. What remains unclear is how suspended elastic shells respond to such flows, particularly when active fluids exist both inside and outside.

In this setting, two active nematic regions act simultaneously on the shell, producing highly nontrivial interactions between internal and external flows. Geometric confinement, curvature, and viscous damping inside the shell organize defects into stable patterns or oscillations~\cite{topo-dynamics-AN-vesicles,topo-morfo-A-shells, Hardoin2022}, while the exterior supports active turbulence and long-range defect dynamics. The resulting mismatch of stresses across the boundary creates a unique playground for emergent dynamics, including spontaneous rotation, symmetry breaking, and shape instabilities~\cite{red-bcell-flicker,defect-dynamics-AN}.

In this letter, we investigate these phenomena by coupling hydrodynamic simulations of active nematics to an elastic-shell model. We find that circular capsules with a suitable size develop persistent rotation, unlike filled disks. Strikingly, boomerang-shaped capsules acquire motion coupled to the capsule orientation. In both cases, the dynamics originate from the coupling between the motion of topological defects and the capsule shape. We further study the role of shell flexibility.

\begin{figure*}[ht]
    \centering
    \includegraphics[width=0.99\linewidth]{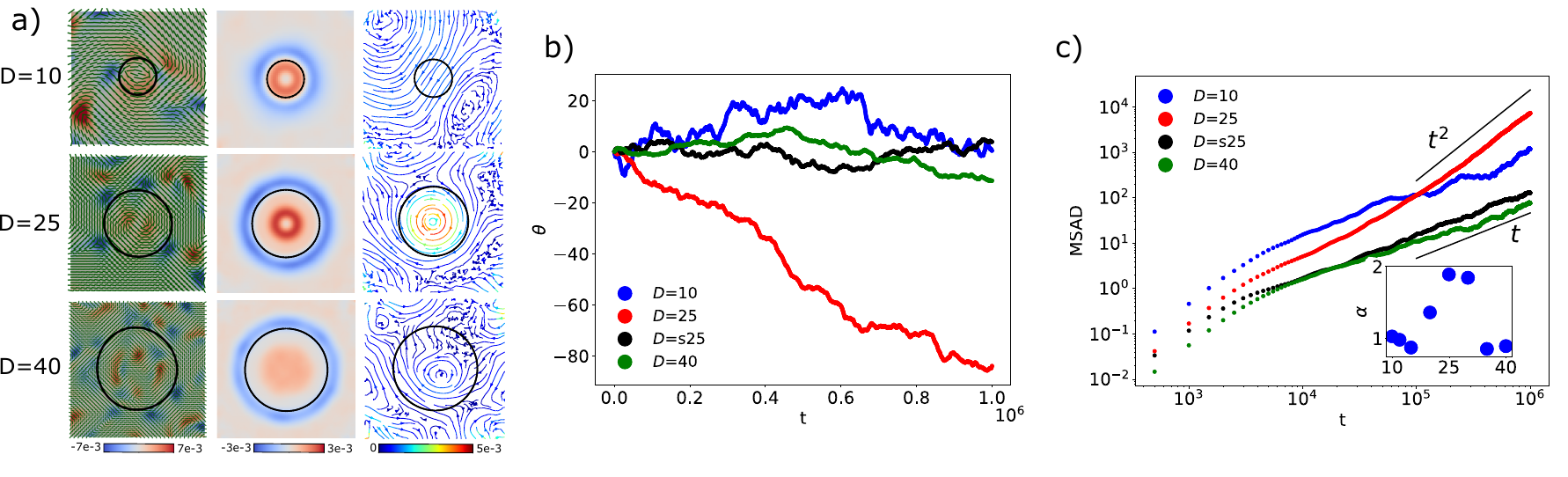}
    \caption{Spontaneous rotation of circular capsules. (a) Fields around capsules of different sizes. From left to right: director field (lines) and charge density (colors) at different times; time-averaged charge density; and streamlines for the time-averaged velocity field, where the color represents its magnitude. The time-averages follow the orientation of the capsule~\cite{SM}. (b) Time evolution of the total angle of rotation for capsules with different sizes. The capsule ``s25'' represents a solid capsule with $D=25$. (c) Mean squared angular displacement versus time for capsules of different sizes. Two slopes are indicated as a reference. The inset indicates the slope of the curves as a function of the diameters. }
    \label{fig1}
\end{figure*}

The active nematic is described within a standard two-dimensional continuum hydrodynamic framework. The local orientational state is represented by the director field $\mathbf{n}$, specifying the mean particle alignment, while the degree of order is quantified by the scalar order parameter $S$. Because particles are apolar, the appropriate description is the symmetric, traceless tensor order parameter, $Q_{\alpha\beta} = 2S \left(n_{\alpha} n_{\beta} - \delta_{\alpha\beta}/2 \right)$, which couples to the flow through the Beris–Edwards equation, together with the incompressibility condition and the Navier–Stokes equation.
~\cite{beris1994thermodynamics, PhysRevLett.89.058101, doostmohammadi2018active}:
\begin{align}
  &\partial _t Q_{\alpha \beta} + v _\gamma \partial _\gamma Q_{\alpha \beta} - S_{\alpha \beta} = \Gamma H_{\alpha\beta}, \label{eq:beris-edwards}\\
  &\partial_\alpha v_\alpha =0, \label{eq:continuity}\\
  &\rho \partial _t v_\alpha + \rho v_\beta \partial_\beta v_{\alpha} = -\partial_\alpha p + 2 \eta \partial_\beta D_{\alpha\beta} - \chi v_\alpha - \zeta \partial_\beta Q_{\alpha\beta}. \label{eq:navier-stokes}
\end{align}
Here $\mathbf{v}$ is the velocity field, $\rho$ the fluid density, $p$ the pressure, $\eta$ the viscosity, $\zeta$ the activity parameter, $\chi$ the friction coefficient, $\Gamma$ the rotational diffusivity, $H_{\alpha\beta}$ the molecular field, and $S_{\alpha\beta}$ the co-rotational term. Capsules are modeled as deformable elastic shells immersed in the active fluid and coupled to the hydrodynamics through an immersed-boundary lattice Boltzmann scheme~\cite{peskinIBM, feng-ib-lbm, Timm-Kruger, D3SM01648J}. At the capsules' boundaries, no-slip boundary conditions and planar anchoring are applied. The results are presented in lattice units, where the lattice spacing between the nodes in the simulations, the time step, and the reference density are set to one. All sample averages considered 120 samples, and the simulation box dimensions are $1024\times 1024$ with periodic boundary conditions. Full details of the model, including free-energy contributions, boundary conditions, and the numerical implementation of the immersed-boundary method, are provided in the Supplementary Material (SM)~\cite{SM}


A striking observed feature of circular capsules is that they can undergo persistent rotation, despite their geometrical symmetry. This effect is only found for capsules of a specific size, which stabilizes a pair of $+1/2$ defects rotating in a yin–yang–like configuration. For smaller diameters, such a defect pair cannot form, while for larger ones additional defects appear, leading to irregular and less predictable dynamics. Importantly, when the capsule is filled (solid), denoted as ``$D=s25$'' in Fig.~\ref{fig1}, persistent rotation is absent, in agreement with recent experimental results~\cite{engrenagem-AN}.

Figure~\ref{fig1}(a) shows the director configuration, topological charge density and flow fields for capsules of different diameters. Only for capsules with $D=25$ are observed two stable and rotating $+1/2$ defects, which generate a coherent circulating flow. This size is consistent with the average distance between defects, $D_{def}=24.6$, which is the characteristic length scale of the unconfined turbulent active nematic~. The associated time-averaged velocity field inside the capsule is enhanced compared to smaller or larger diameters, consistent with the persistence of the rotation. This effect is also observed in the time-averaged spatial distribution of the topological charge density. In two dimensions, the charge density can be written in terms of the $Q$-tensor as $q(\mathbf r) \;=\; \frac{1}{2\pi}\,
\Big(\partial_x Q_{xx}\,\partial_y Q_{xy}
-\partial_x Q_{xy}\,\partial_y Q_{xx}\Big)$. For $D=25$, the time-averaged charge density is localized around a circle concentric with the capsule boundaries due to the closed trajectories of the two rotating defects. This behavior is consistent with experiments on active nematics confined in circular cavities, where pairs of defects rotate persistently~\cite{capsulas-exp,rotating-defects}. In our case, the boundaries are free to move and thus the capsules undergo global rotation as a consequence of angular momentum conservation. Importantly, the internal defect dynamics can drive rotation of the capsule but cannot generate net translation. This distinction will play a central role in our discussion of boomerang-shaped capsules. For circular capsules, the center of mass exhibits diffusive motion, with a diffusion coefficient that decreases with increasing capsule diameter, as shown in Fig.~S7.

\begin{figure*}
    \centering
    \includegraphics[width=0.99\linewidth]{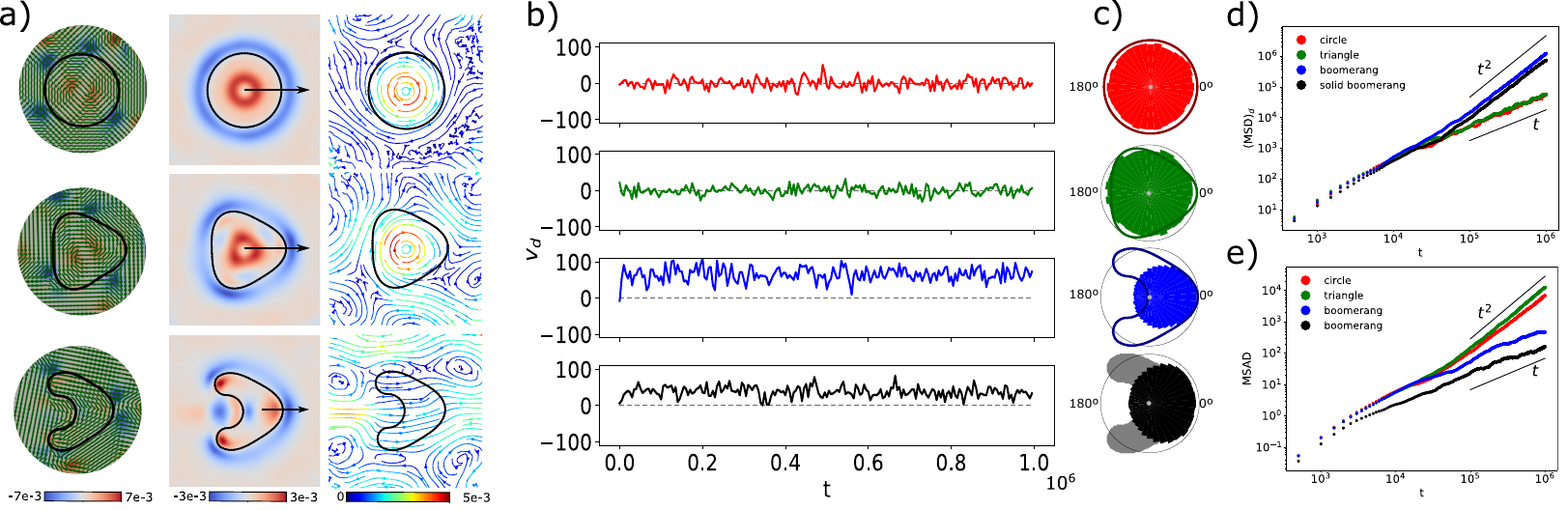}
    \caption{Directed motion of capsules with different shapes along their axis of symmetry. (a) Fields around capsules of different shapes. From left to right: director field (lines) and charge density (colors) at different times; time-averaged charge density; and streamlines for the time-averaged velocity field, where the color represents its magnitude. The averages follow the capsule orientation~\cite{SM}. The arrows indicate the symmetry axis of the capsules, $\mathbf{e}_d$. (b) Capsule velocity along its symmetry axis for the different shapes indicated in ``c''. The last one is for a solid capsule with boomerang shape. The dashed lines indicates the zero. (c) Histogram of the angle between the velocity field and the symmetry axis for the different shapes: circle, triangle, boomerang and solid boomerang. (d) Mean squared displacement in the direction of the symmetry axis ($MDS_d$) and mean squared angular displacement (MSAD) versus time for different shapes. Two slopes in each plot are indicated as a reference.}
    \label{fig2}
\end{figure*}

The rotational dynamics are quantified in Fig.~\ref{fig1}(b), where the total angle of rotation of the capsule is shown as a function of time. Only for $D=25$ is observed a steady and monotonic increase in the angular distance, corresponding to persistent rotation. For all other cases, including the solid capsule with $D=25$ the angle fluctuates randomly without a clear trend. 

A more detailed characterization is obtained from the mean-squared angular displacement (MSAD), calculated as $\mathrm{MSAD}(t) = \left\langle \left[ \theta(t_0+t) - \theta(t_0) \right]^2 \right\rangle_{t_0}$,
where $\theta(t)$ is the angle of rotation of the capsule and $\langle \cdot \rangle_{t_0}$ denotes a time origin average. Figure~\ref{fig1}(c) reveals that the MSAD grows linearly in time, with slope close to $2$, only for $D=25$. By contrast, for other diameters, as well as for the solid capsule with $D=25$, the slope is close to $1$, consistent with random rotational diffusion. The inset of Fig.~\ref{fig1}(c) highlights that the slope varies non-monotonically with the diameter, reaching a maximum close to $2$ at $D=25$.

Taken together, these results demonstrate that persistent rotation of circular capsules emerges only when confinement accommodates precisely two $+1/2$ defects in a yin–yang configuration. The phenomenon is absent in both smaller and larger capsules, and suppressed entirely in solid inclusions, underscoring the essential role of defect–flow coupling in producing emergent motility in active nematic shells. This system demonstrates that it is possible to achieve persistent rotation with a symmetrical structure, which can be used in applications involving micromotors and energy production at the micro scale.


To assess the role of geometry, we simulated capsules with three different shapes and the same size equal to 25: circular, triangular with smooth edges, and boomerang-shaped, see Fig.~\ref{fig2}. The circular and triangular capsules both accommodate a pair of internal defects that rotate persistently, leading to coherent global rotation. This is clearly visible in Fig.~\ref{fig2}(a), which shows the time averaged velocity fields and topological charge densities, and in Fig.~\ref{fig2}(e), where the mean-squared angular displacement (MSAD) exhibits a slope close to $2$. These results are consistent with our previous analysis of circular capsules, confirming that internal defect dynamics drive persistent rotation for appropriate sizes.

In contrast, boomerang-shaped capsules display a qualitatively distinct behavior. They do not rotate persistently: the MSAD in Fig.~\ref{fig2}(e) has slope close to $1$, characteristic of random rotations. Instead, these capsules undergo directed translation along their symmetry axis, draw in Fig.~\ref{fig2}(a). This is evident in Fig.~\ref{fig2}(b), which shows the velocity component along the symmetry axis (always positive for boomerangs, but fluctuating around zero for circles and triangles) and in Fig.~\ref{fig2}(d), where the mean-squared displacement along the symmetry axis (MDS$_d$) grows with slope close to $2$ for boomerangs, while remaining close to $1$ for the other shapes. The mean-squared displacement along the symmetry axis is defined as $\text{MSD}_{d}(t)=\left< [ \int_{0}^{t} (\mathbf{v}(s)\cdot \mathbf{e}_{d}(s)) ds ]^{2} \right>$, where $(\mathbf{v}(t))\cdot \mathbf{e}_{d}(t))$ is the velocity along the capsule's symmetry axis, $\mathbf{e}_{d}$, at time $t$. The integral results in the ``net-distance'' traveled along the symmetry axis.

The mechanism underlying this behavior is again defect-mediated, but with a crucial difference: in boomerangs, internal defects are not stabilized by the capsule shape and size (Fig.~\ref{fig2}a), so there is no persistent internal defect motion to drive rotation. Instead, the U-shaped concavity at the rear of the boomerang favors the trapping of positive defects in the surrounding active nematic. Indeed, a region of enhanced positive charge density is observed behind the capsule in Fig.~\ref{fig2}(a). These external defects, together with the geometric asymmetry of the capsule and the planar anchoring at its surface, generate a net active force that propels the capsule forward. 

A simple estimate illustrates the mechanism. For planar anchoring and constant order parameter $S$, the active force density on a curved boundary is $f_i=-2\zeta S\,\kappa\,n_i$, where $\kappa=1/R$ is the local curvature and $\mathbf n$ the normal. Integrating over a circular arc of opening angle $\Delta\theta$ gives $F \;=\; 4\,\zeta S \,\sin\!\Big(\tfrac{\Delta\theta}{2}\Big)$.
For a semicircular rear ($\Delta\theta=\pi$) this reduces to $F=4\zeta S$, showing that the propulsive force is set by the activity and nematic order, independent of the capsule size for this idealized geometry~\cite{SM}. Note that the contribution from the inner curvature does not generate net translation due to momentum conservation. In addition, the forces associated with the three edges tend to cancel one another, as occurs for the symmetric triangular capsule (Fig.~\ref{fig2}).

As emphasized earlier, internal active fluid can only drive rotation, not translation; here, it is the interaction with the external fluid that is essential for translation-orientation coupling. We note that while the MDS$_d$ along the symmetry axis is ballistic for boomerangs, the overall mean-squared displacement of the center of mass remains diffusive (see Fig. S11).

To further assess the role of internal defects, we also simulated a solid (filled) boomerang-shaped capsule. Its qualitative behavior was indistinguishable from that of the fluid filled boomerang (Fig.~\ref{fig2}), exhibiting directed propulsion along its symmetry axis and only random rotations. This result demonstrates that internal defects play only a minor role in the translational dynamics of boomerang-shaped capsules. Instead, the directed motion arises predominantly from the interaction between the capsule geometry, surface anchoring, and the external defect field.


We now address the role of capsule flexibility. So far, the capsules were quite rigid with only small deformations (Fig.~S6). In recent experiments~\cite{rodrigo-pumps, velez2024}, solid structures were designed with either rigid or flexible backbones, although they were attached to the substrate. Future experiments may consider suspended flexible capsules, for which the effect of deformability becomes particularly relevant. 

\begin{figure}
    \centering
    \includegraphics[width=0.99\linewidth]{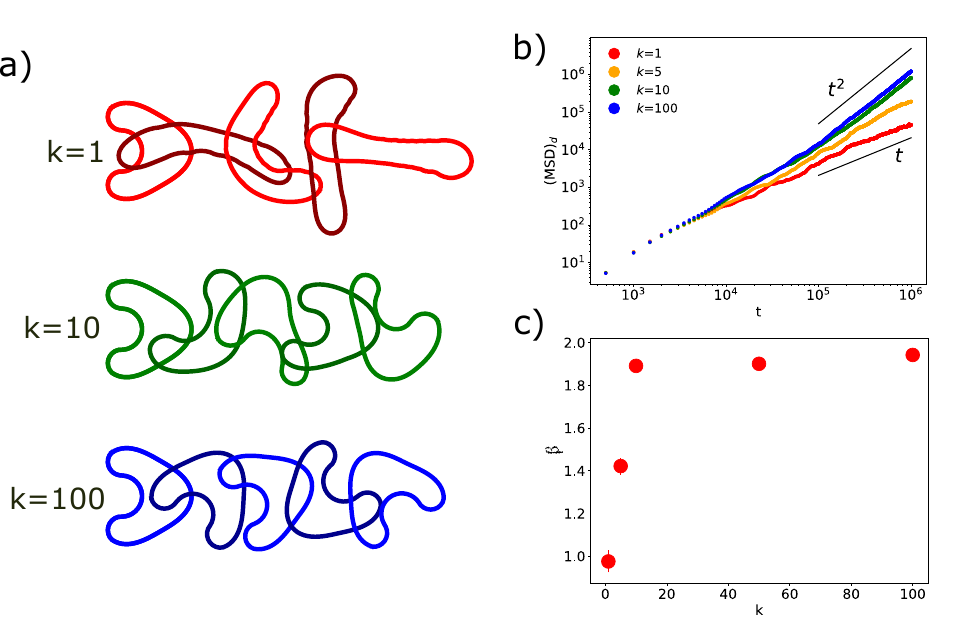}
    \caption{Flexible boomerang-shaped capsules of size $D=25$. (a) Time evolution of the shape of capsules with different stiffness. (b) Time evolution of the mean squared displacement along the capsule symmetry axis for different values of stiffness. Two slopes are indicated as reference. (c) Slopes of the curves as a function of the stiffness.}
    \label{fig3}
\end{figure}

\begin{figure}
    \centering
    \includegraphics[width=0.99\linewidth]{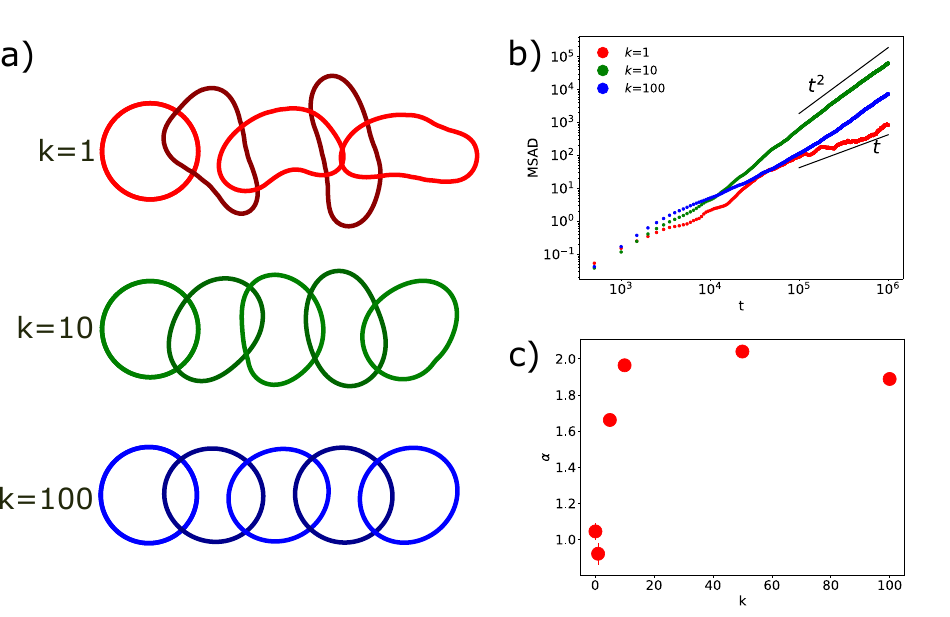}
    \caption{Flexible circular capsules of size $D=25$. (a) Time evolution of the shape of capsules with different stiffness. (b) Time evolution of the mean squared angular displacement for different values of stiffness. Two slopes are indicated as reference. (c) Slopes of the curves as a function of the stiffness.  }
    \label{fig4}
\end{figure}

In our simulations, capsule rigidity is controlled by the parameter $k$, which tunes both bending and stretching resistance ~\cite{SM}. Figures~\ref{fig3} and~\ref{fig4} indicate that a minimum rigidity is required for capsules to sustain persistent rotation or translation–orientation coupling. For boomerang-shaped capsules (Fig.~\ref{fig3}(a)), when $k<10$ the capsule cannot maintain its geometry, losing the U-shaped rear that accommodates the positive defects. As a consequence, no coherent propulsion is observed. This is reflected in the mean-squared displacement along the symmetry axis (MDS$_d$): for $k>10$ the slope is close to $2$, indicating ballistic translation aligned with the capsule axis, while for $k<10$ the slope is close to $1$, corresponding to random motion (Fig.~\ref{fig3}(b)). The slopes as a function of rigidity (Fig.~\ref{fig3}(c)) highlight a clear transition around $k\simeq 10$.

A similar effect occurs for circular capsules, Fig.~\ref{fig4}. For low rigidities, the capsule shape deforms substantially, stretching along one direction and thus preventing the formation of a stable rotating pair of $+1/2$ defects. As a result, the dynamics reduce to random rotations, with the MSAD displaying a slope close to $1$, Fig.~\ref{fig4}(b). Above $k\simeq 10$, the capsules preserve their circular geometry and accommodate the yin–yang defect pair, recovering persistent rotation with MSAD slope close to $2$ (Fig.~\ref{fig4}(c)).

This rigidity threshold can be rationalized in terms of a nondimensional ratio comparing elastic and active stresses. The characteristic active stress scales as $\sigma_\text{a}\sim \zeta$, while the elastic stress resisting deformations scales as $\sigma_\text{e}\sim k/R^2$, with $R$ the capsule radius. Persistent rotation or propulsion is only sustained when $\sigma_\text{e}/\sigma_\text{a}\gtrsim 1$, corresponding to $k \gtrsim k_c \sim \zeta R^2 \simeq 6.3$. The observed transition around $k\simeq 10$ is consistent with this scaling estimate, indicating that rigidity must be sufficient to suppress active-flow–induced distortions that otherwise destroy the defect structures responsible for coherent motion.

In all cases, the center-of-mass motion of both flexible and rigid capsules remains diffusive (see Figs.~S9-S11). However, the effective diffusion coefficient decreases with decreasing rigidity, as flexible capsules dissipate more energy into shell deformations rather than transmitting it to translational motion in the active fluid.


The existence of a rigidity-controlled transition also points to potential applications in drug delivery and microfluidics. For example, rigid boomerang-shaped capsules could be engineered to exploit translation–orientation coupling for guided transport. Upon reaching a target environment, their rigidity could be reduced, e.g. through temperature or chemical triggers, causing them to lose directional propulsion and switch to random motion, thereby enhancing local release and mixing of encapsulated compounds.

In summary, we have shown that suspended elastic capsules in active nematics display a rich variety of defect-mediated dynamics governed by their size, shape, and rigidity. Circular capsules can rotate persistently despite their geometrical symmetry, provided their diameter stabilizes a yin–yang pair of defects, while boomerang-shaped capsules break symmetry and drive translation-orientation coupling. Flexibility plays a central role: below a critical rigidity, capsules deform, fail to sustain stable defect structures, and lose persistent motion. These results highlight defect–capsule coupling as a general mechanism for emergent motility in active fluids and suggest design principles for microswimmers and active delivery vehicles, where shape and mechanical properties may be tuned to control transport and release.

We acknowledge financial support from the Portuguese Foundation for Science and Technology (FCT) under the contracts: UIDB/00618/2020 (DOI: 10.54499/UIDB/00618/2020),  UIDP/00618/2020 (DOI: 10.54499/UIDP/00618/2020) and 2023.10412.CPCA.A2 (DOI: 10.54499/2023.10412.CPCA.A2).

\bibliography{references}
\bibliographystyle{apsrev4-2}

\end{document}